\documentclass[3p,times,twocolumn]{elsarticle}

\usepackage{ecrc}

\volume{00}

\firstpage{1}

\journalname{Nuclear Physics B Proceedings Supplement}

\runauth{}

\jid{nuphbp}

\jnltitlelogo{Nuclear Physics B Proceedings Supplement}

\usepackage{amssymb}

\usepackage[figuresright]{rotating}

\begin{document}

\begin{frontmatter}



\dochead{}

\title{The Physics Landscape after the Higgs Discovery at the LHC}


\author{John Ellis}

\address{Theoretical Particle Physics and Cosmology Group, Physics Department, King's College London, London WC2R 2LS, UK; \\
TH Division, Physics Department, CERN, CH-1211 Geneva 23, Switzerland \\
~~\\
{\tt KCL-PH-TH/2015-19, LCTS/2015-09, CERN-PH-TH/2015-085}
}

\begin{abstract}
What is the Higgs boson telling us? What else is there, maybe supersymmetry and/or dark matter? How do we find it?
These are now the big questions in collider physics that I discuss in this talk, from a personal point of view.
\end{abstract}


\end{frontmatter}



\section{Introduction}

\begin{figure}[htb]
\centering
\includegraphics[height=2.1in]{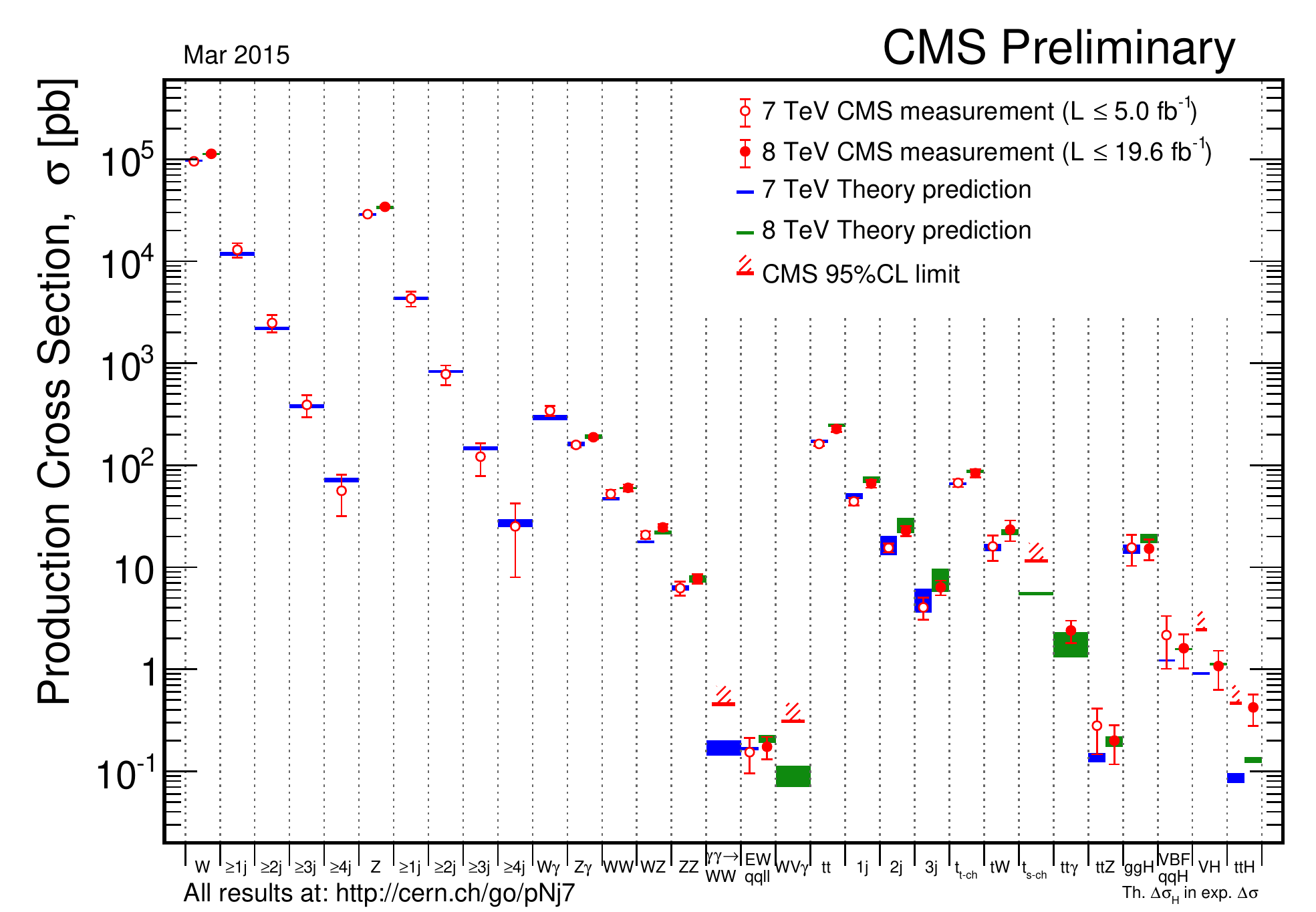}
\caption{\it A compilation of cross sections at the LHC measured by the CMS Collaboration~\protect\cite{CMSxs}.}
\label{fig:sections}
\end{figure}

The Standard Model (SM) has passed the tests provided by Run 1 of the LHC with flying colours.
Many cross sections for particle and jet production have been measured at the LHC~\cite{CMSxs}, 
and are in agreement with the SM predictions, as seen in Fig.~\ref{fig:sections}. Jet 
production cross sections agree over large ranges in energy and many orders of magnitude with QCD
calculations within the SM, as do
measurements of single and multiple $W^\pm$ and $Z^0$ production and
measurements of single and pair production of the top quark. The biggest headline of LHC Run 1
was of course the discovery by CMS and ATLAS of a (the?) Higgs boson~\cite{Higgs},
which has now been detected in three production channels, as also seen in Fig.~\ref{fig:sections}, also
in agreement with the SM predictions. Much of this talk will concern what we already know about this newly-discovered particle,
and the hints it may provide for other new physics, as well as other topics within and beyond the Standard Model.

\section{QCD}

\begin{figure}[htb]
\centering
\includegraphics[height=2in]{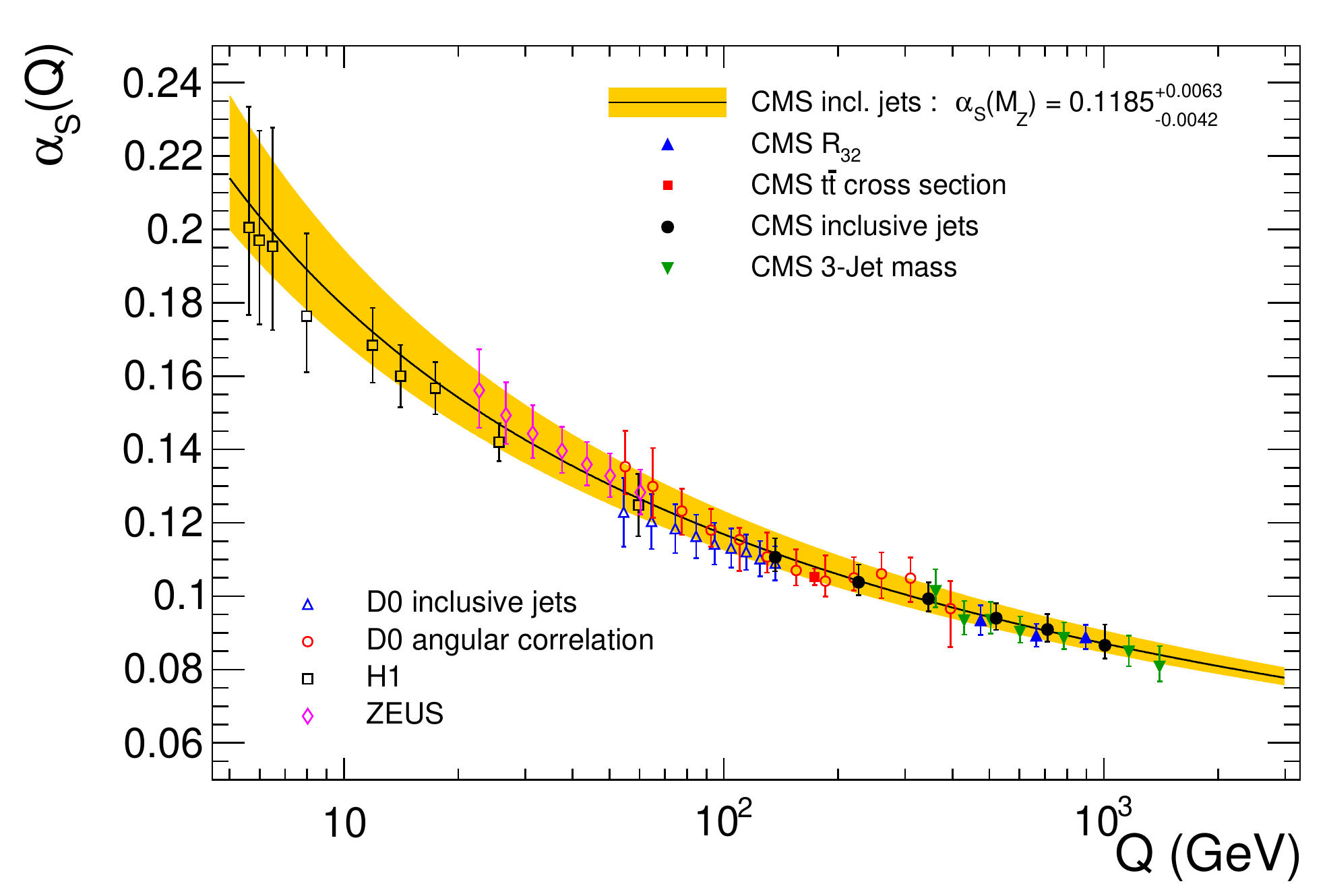}
\caption{\it Jet production measurements at the LHC show that $\alpha_s$ continues to run downward at
energies beyond $1$~TeV~\protect\cite{CMSalphas}.}
\label{fig:run}
\end{figure}

QCD is the basis for LHC physics: it provides many tests of the Standard Model
as well as dominating particle production and deluging us with
with backgrounds and pile-up events. The agreement between QCD predictions and measurements of large-$p_T$ jet production
at the LHC over many orders of magnitude yields measurements of
the strong coupling that are consistent with the world average value $\alpha_s (M_Z) = 0.1185 \pm 0.0006$,
and demonstrate that $\alpha_s$ continues to run downward beyond the TeV
scale~\cite{CMSalphas}, perhaps towards grand unification, as seen in Fig.~\ref{fig:run}.

Not only are perturbative QCD calculations doing a fantastic job overall of predicting the production cross sections
for jets and massive vector bosons, but also for the Higgs boson. Accurate higher-order QCD calculations are at a premium
for the dominant gluon-fusion contribution to the Higgs production cross section. Several
different NNLO calculations are available, and are included in various publicly-available
tools~\cite{HiggsxsWG}. Unfortunately, the agreement between them is not yet perfect.
Fortunately, progress is being made on NNNLO calculations~\cite{NNNLO}.
These will improve the theoretical accuracy, but progress in convergence between the
parton distribution functions will also be needed in order to reduce the theoretical uncertainties below the
experimental measurement uncertainties.

\section{Flavour Physics}

\begin{figure}[htb]
\centering
\includegraphics[height=2.3in]{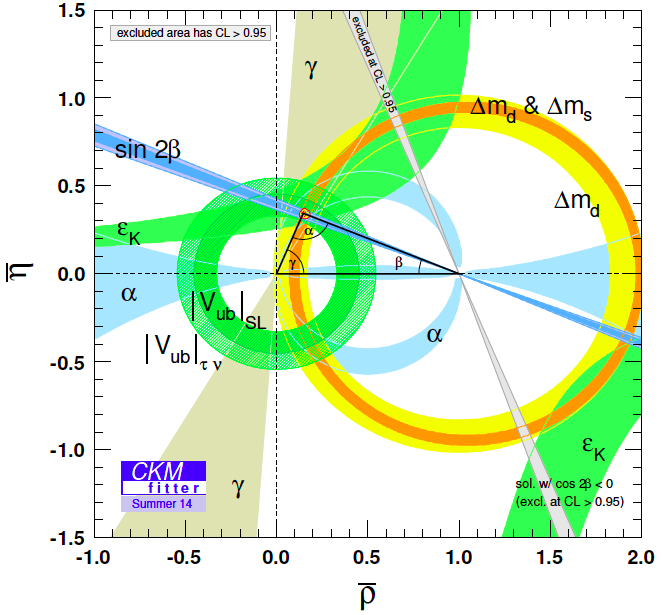}
\caption{\it Flavour and CP violation measurements generally agree well with the CKM paradigm~\protect\cite{CKMFitter}.}
\label{fig:hs}
\end{figure}

\begin{figure}[htb]
\centering
\includegraphics[height=1.3in]{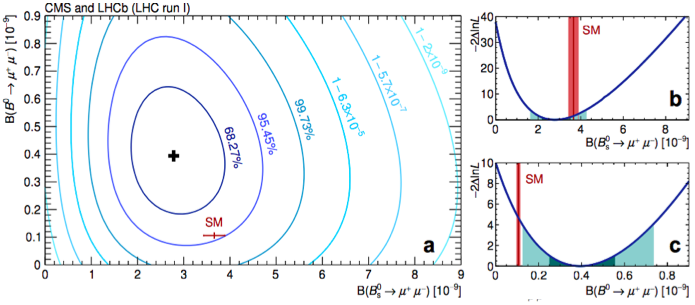}
\caption{\it {\bf a}: Measurements by the CMS and LHCb Collaborations
of $B_{s,d} \to \mu^+ \mu^-$ decays, including {\bf b} a
clear signal for $B_s \to \mu^+ \mu^-$ decay that agrees with the SM, and {\bf c} a hint of $B_d \to \mu^+ \mu^-$ decay,
possibly at a rate larger than expected in the SM~\protect\cite{Bsmumu}.}
\label{fig:Bmumu}
\end{figure}

Another pillar of the SM is the Cabibbo-Kobayashi-Maskawa (CKM) model of flavour mixing and CP violation.
It is in general very successful, as seen in Fig.~\ref{fig:hs}~\cite{CKMFitter}.
For example, the second-greatest discovery during Run~1 of the LHC
was perhaps the measurement by the CMS and LHCb Collaborations of the rare decay $B_s \to \mu^+ \mu^-$,
with a branching ratio in good agreement with the SM prediction~\cite{Bsmumu}:
\begin{equation}
BR(B_s \to \mu^+ \mu^-) \; = \; 2.8^{+0.7}_{-0.6} \times 10^{-9} \, ,
\label{Bsmumu}
\end{equation}
as seen in Fig.~\ref{fig:Bmumu}. However, the joint CMS and LHCb analysis~\cite{Bsmumu}
also has an suggestion of a $B_d \to \mu^+ \mu^-$ signal
that is larger than the SM prediction:
\begin{equation}
BR(B_d \to \mu^+ \mu^-) \; = \; 3.9^{+1.6}_{-1.4} \times 10^{-10} \, ,
\label{Bdmumu}
\end{equation}
as also seen in Fig.~\ref{fig:Bmumu}. If confirmed, this measurement would conflict not just with the SM,
but also models with minimal flavour violation (MFV), including many supersymmetric scenarios.
Something to watch during Run~2!

There is scope elsewhere for deviations from CKM predictions:
for example, the data allow an important contribution to $B_s$ meson mixing
from physics beyond the SM (BSM)~\cite{CKMFitter}.
Also, there are issues with $e - \mu$ universality in semileptonic $B$ decays~\cite{nonuniv}
and a persistent anomaly in the $P_5^\prime$ angular distribution for
$B^0 \to K^{*0} \mu^+ \mu^-$~\cite{P5prime}. Could this be related to the intriguing excess in $H \to \mu \tau$
decay reported by the CMS Collaboration~\cite{CMSHmutau}, which is discussed later?
Other points to watch include discrepancies in the determinations of the
$V_{ub}$ CKM matrix element and the Tevatron
diimuon asymmetry anomaly~\cite{dimuon}.
However, some anomalies do seem to be going away, such as the branching ratio for
$B_u \to \tau^+ \nu$ decay, which is now in good agreement with the SM~\cite{Bellebtn} and the forward-backward asymmetry in
$t \bar{t}$ production~\cite{ttbarAFB}, which is consistent with the latest higher-order QCD calculations~\cite{Brodsky}, as is the
 $t \bar{t}$ rapidity asymmetry measured at the LHC. However, there are still plenty of
 flavour physics issues to be addressed during LHC Run~2.

\section{Higgs Physics}

The Higgs boson may be regarded as, on the one hand, the capstone of the glorious arch of the SM or,
on the other hand, as the portal giving access to new physics. In this Section we discuss first the extent to which the new particle
discovered on July 4th, 2012 fulfils its SM r\^ole, and then what hints it may be able to provide about possible BSM physics.

\subsection{Mass Measurements}

\begin{figure}[htb]
\centering
\includegraphics[height=2in]{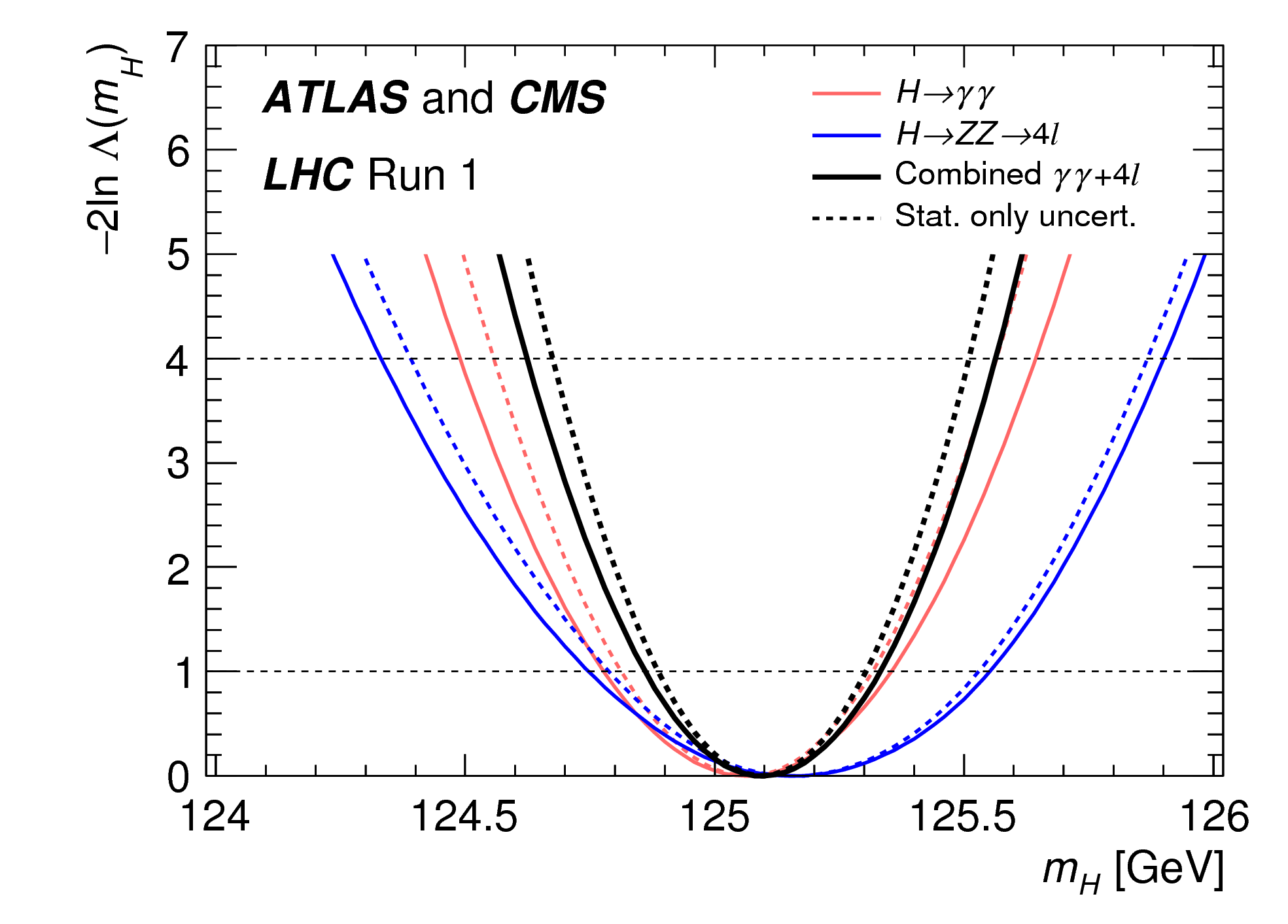}
\caption{\it Measurements of $m_H$ by ATLAS and CMS in the $\gamma \gamma$ and $ZZ^* \to 2 \ell^+ 2 \ell^-$ final states,
as complied in~\protect\cite{MHjoint}.}
\label{fig:MH}
\end{figure}

The mass of the Higgs boson is measured most accurately
in the $\gamma \gamma$ and $Z Z^* \to 2 \ell^+ 2 \ell^-$ final states, and ATLAS and CMS have both
reported accurate measurements in each of these final states as shown in Fig.~\ref{fig:MH}. ATLAS measures~\cite{ATLASmH}
\begin{eqnarray}
H \to \gamma \gamma: m_H & = \; 126.02 \pm 0.51~{\rm GeV}\, , \nonumber \\
H \to Z Z^*: m_H & = \; 124.51 \pm 0.52~{\rm GeV}\, ,
\label{ATLASm}
\end{eqnarray}
and CMS measures~\cite{CMSmH}
\begin{eqnarray}
H \to \gamma \gamma: m_H & = \; 124.70 \pm 0.34~{\rm GeV}\, , \nonumber \\
H \to Z Z^*: m_H & = \; 125.59 \pm 0.45~{\rm GeV} \, .
\label{CMSm}
\end{eqnarray}
Combining all these measurements, the ATLAS and CMS Collaborations find~\cite{MHjoint}
\begin{equation}
m_H \; = \; 125.09 \pm 0.24~{\rm GeV} \, .
\label{mH}
\end{equation}
In addition to being a fundamental measurement in its own right, and casting light on the possible validity of various BSM models
(for example, this value is perfectly consistent with supersymmetric predictions~\cite{SUSYmH}),
the precise value of $m_H$ is also important for the stability of the electroweak vacuum in the Standard Model~\cite{Buttazzo},
as discussed later.

\subsection{The Higgs Spin and Parity}

\begin{figure}[htb]
\centering
\includegraphics[height=2in]{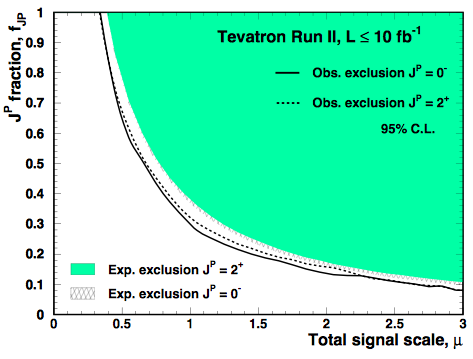}
\caption{\it Tests of spin-parity hypotheses for the Higgs boson via the kinematics of associated $H + W^\pm/Z^0$ production
at the Tevatron~\protect\cite{TevatronHspin}.}
\label{fig:Higgsspin}
\end{figure}

The fact that the Higgs boson decays into $\gamma \gamma$ excludes spin 1,
and spin 0 is expected, but spins 2 and higher are also possible in principle. The Higgs spin has been probed in many
ways~\cite{ATLASHspin,CMSHspin,TevatronHspin},
via its production and decay rates~\cite{ESYrates}, the kinematics of Higgs production
in association with the $W^\pm$ and $Z^0$~\cite{EHSY}, and decay angular distributions for $W^+ W^-$,
$ZZ$ and $\gamma \gamma$ final states~\cite{EFHSY}. The results of tests using the kinematics of associated
$H + W^\pm/Z^0$ production at the Tevatron are shown in Fig.~\ref{fig:Higgsspin}~\cite{TevatronHspin}.
By now there is overwhelming evidence against the Higgs boson having spin 2. Moreover, as also seen in
Fig.~\ref{fig:Higgsspin}~\cite{TevatronHspin}, it has been established that its couplings to
$W^+ W^-$ and $ZZ$ are predominantly CP-even, i.e., it couples mainly as a scalar, not as a pseudoscalar.

\subsection{Higgs Couplings}

\begin{figure}[htb]
\centering
\includegraphics[height=2.3in]{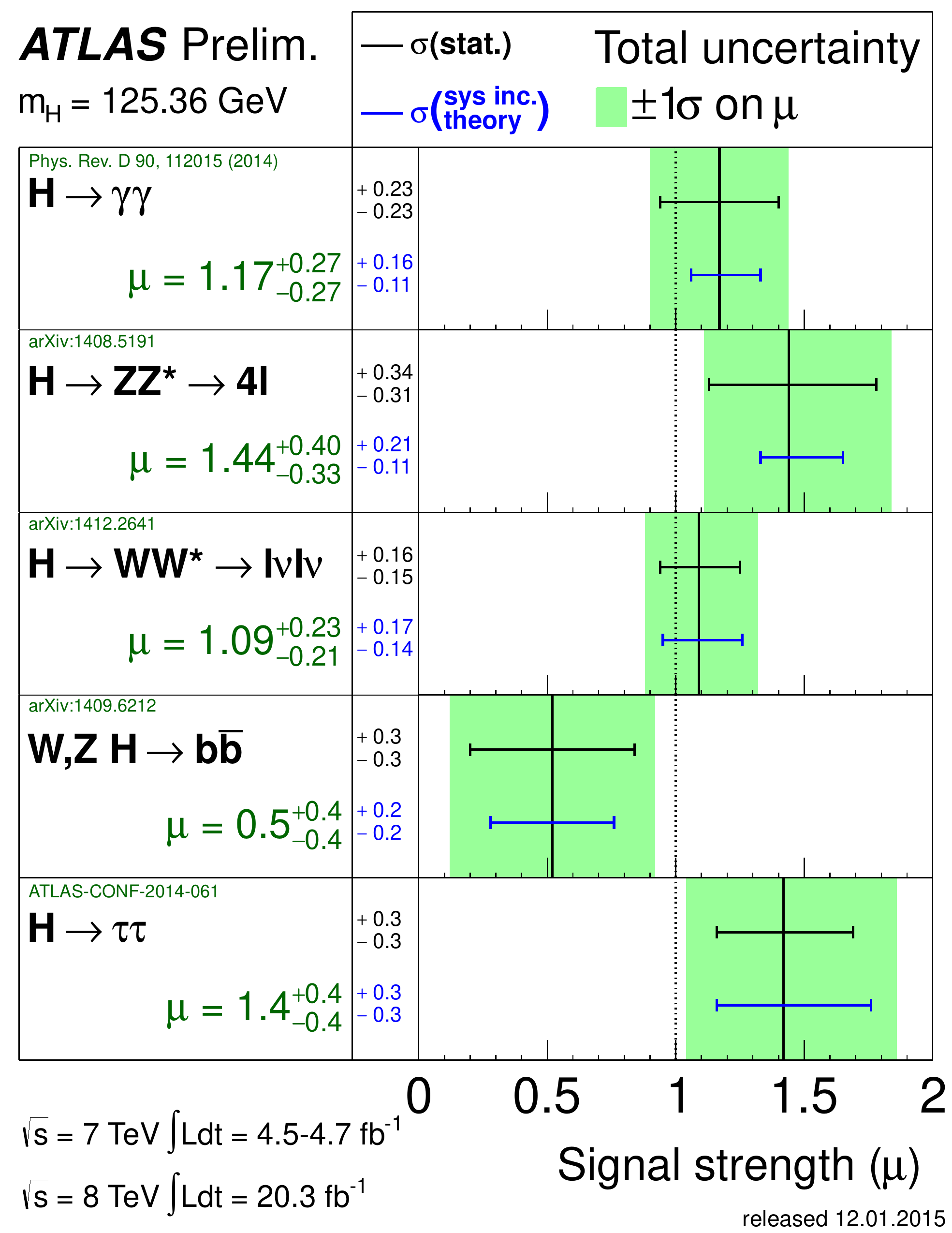}
\caption{\it The Higgs signal strengths $\mu$, normalised to unity for the SM, as measured by
ATLAS~\protect\cite{ATLASmu}.}
\label{fig:strengths}
\end{figure}

As seen in Fig.~\ref{fig:strengths}, the strengths of the Higgs signals measured
by ATLAS in individual channels~\cite{ATLASmu} are generally
compatible with the SM predictions (as are CMS measurements~\cite{CMSmu}) within the statistical fluctuations,
which are inevitably large at this stage. ATLAS and CMS report the following overall signal strengths after combining their measurements
in the $\gamma \gamma$,  $Z Z^*$, $W W^*$, $b \bar{b}$ and $\tau^+ \tau^-$ channels:
\begin{eqnarray}
{\rm ATLAS:} ~\mu & = \; 1.30 \pm 0.12 \pm 0.10 \pm 0.09 \, , \nonumber \\
{\rm CMS:} ~ \mu & = \; 1.00 \pm 0.09~^{+ 0.08}_{- 0.07} \pm 0.07 \, .
\label{mu}
\end{eqnarray}
Both averages are quite compatible with the SM and with each other,
as is also true of the Tevatron measurements~\cite{TevatronH}.

\begin{figure}[htb]
\centering
\includegraphics[height=2.15in]{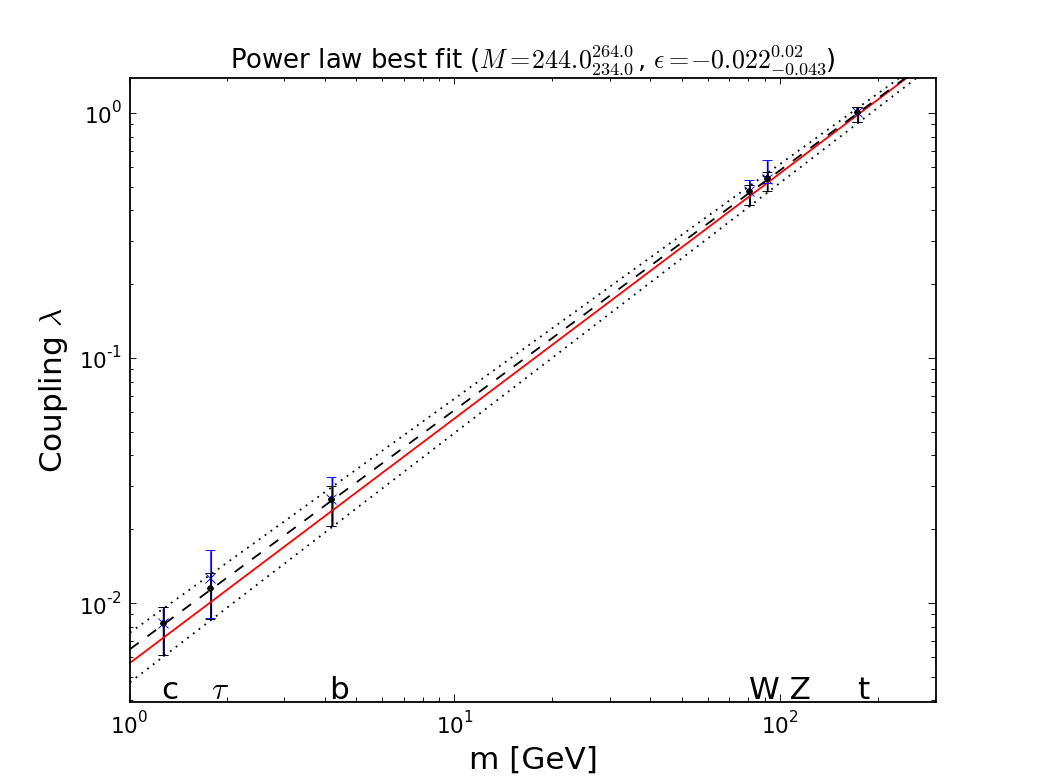}
\caption{\it A global fit to the $H$ couplings of the form (\protect\ref{epsilon}) (central values as dashed and
$\pm$1$\sigma$ values as dotted lines), which is very consistent
with the linear mass dependence for fermions and quadratic mass dependence for bosons (solid
red line) expected in the SM~\protect\cite{EY3}.}
\label{fig:Mass_dependence}
\end{figure}

Because of its connection to mass generation, the couplings of the Higgs boson
to other particles in the SM should be related to their masses:
linearly for fermions, quadratically for bosons, and be scaled by the Higgs vev $v = 246$~GeV.
These predictions are implicit in the
measurements in Fig.~\ref{fig:strengths}, and are tested directly
in Fig.~\ref{fig:Mass_dependence}. The latter displays a global fit in which the Higgs coupling data are
parametrised as~\cite{EY3}
\begin{equation}
\lambda_f \; = \; \sqrt{2} \left( \frac{m_f}{M} \right)^{(1 + \epsilon)}, \; \; 
g_V \; = \; 2 \left( \frac{M_V^{2(1 + \epsilon)}}{M^{(1 + \epsilon)}} \right) \, .
\label{epsilon}
\end{equation}
As seen in the left panel of Fig.~\ref{fig:Mass_dependence}, the data yield
\begin{equation}
\epsilon \; = \; - 0.022^{+0.020}_{-0.043}, \; \; M \; = \; 244^{+20}_{-10}~{\rm GeV}, \,
\label{epsilonM}
\end{equation}
which is in excellent agreement with the SM predictions $\epsilon = 0$, $M = 246$~GeV.
Similar results have also been found recently in an analysis by the CMS Collaboration~\cite{CMSmu}.
It seems that Higgs couplings indeed have the expected characteristic dependence
on particle masses.

According to the SM, flavour should be conserved to a very good
approximation in Higgs couplings to fermions. This is consistent with the available upper limits on low-energy effective
flavour-changing interactions, which would, however, also allow lepton-flavour-violating
Higgs couplings that are much larger than the SM predictions~\cite{BEI}. Looking for such
interactions is therefore a possible window on BSM physics. On the basis of low-energy data, we estimated that
the branching ratios for $H \to \mu \tau$ and $H \to e \tau$ decays could each be
as large as ${\cal O}(10)$\%, i.e., as large as BR$(H \to \tau \tau$, whereas the branching ratio for $H \to \mu e$
could only be much smaller, $\lesssim 10^{-5}$~\cite{BEI}. The CMS Collaboration has recently reported
a measurement~\cite{CMSHmutau}
\begin{equation}
{\rm BR}(H \to \mu \tau) \; = \; 0.89^{+0.40}_{-0.37} \, \% \, ,
\label{Hmutau}
\end{equation}
which is $\sim 2.5 \sigma$ different from zero.
SM flavour physics predictions are therefore being probed more stringently by the LHC than by low-energy experiments, and we are keen
to see corresponding results from ATLAS and from Run~2 of the LHC!

\begin{figure}[htb]
\centering
\includegraphics[height=3in]{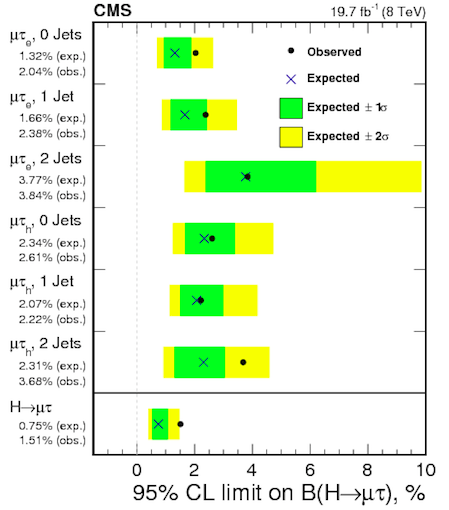}
\caption{\it Results from the CMS search for $H \to \mu \tau$ decay~\protect\cite{CMSHmutau}.}
\label{fig:Hmutau}
\end{figure}

\begin{figure}[htb]
\centering
\includegraphics[height=2.5in]{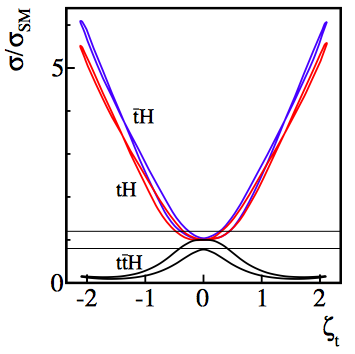}
\caption{\it The effects of a CP-violating coupling on the $H t {\bar t}$, $H t$ and $H {\bar t}$ production
cross sections, taking into account the current constraints from the $Hgg$ and $H \gamma \gamma$ couplings~\protect\cite{EHST}.}
\label{fig:tHCP}
\end{figure}

Although all the indications are that the dominant Higgs couplings are CP-even, as seen, e.g., in Fig.~\ref{fig:Higgsspin} above,
there may also be an admixture of CP-odd couplings, 
whose fraction may depend on the particle whose coupling to the Higgs boson are
being probed. Since the leading CP-odd $H$ coupling to fermions would have the
{\it same} (zero) dimension as the leading CP-even coupling, whereas the leading CP-odd $H$ coupling to
massive vector bosons would have {\it higher} dimension than the leading CP-even coupling,
the latter may be more suppressed. Ideas for probing CP violation in $H \to \tau^+ \tau^-$ decay
have been suggested~\cite{Askewetal}, and CP violation may also be probed in the $H t {\bar t}$
couplings~\cite{EHST}. As seen in Fig.~\ref{fig:tHCP}, this could affect the total cross sections for associated $H t {\bar t}$,
$H t$ and $H {\bar t}$ production, shown as
functions of $\zeta_t \equiv$ arctan(CP-odd coupling/CP-even coupling). If $\zeta_t \ne 0$,
a CP-violating transverse polarization asymmetry is in principle observable in $H t$ and $H {\bar t}$ production,
as discussed in~\cite{EHST}.

\subsection{Is the Higgs Boson Elementary or Composite?}

One of the key questions about the Higgs boson is whether
it is elementary or composite. One might have thought that a composite Higgs boson would
naturally have a mass comparable to the scale of compositeness, but the mass can be
suppressed if it is a pseudo-Nambu-Goldstone boson with a mass that is protected by some approximate symmetry,
perhaps becoming consistent with the measured Higgs mass $\sim 125$~GeV.
This possibility may be probed using a phenomenological
Lagrangian ${\cal L}$ with free $H$ couplings, that may be
constrained using $H$ decay and production data.
Since the Standard Model relation $\rho \equiv m_W/m_Z \cos \theta_W = 1$ agrees well with the data,
one usually assumes that the phenomenological Lagrangian has a custodial symmetry: SU(2)$\times$SU(2) $\to$ SU(2). 
Then one may parametrise the leading-order terms in ${\cal L}$ as follows:
\begin{eqnarray}
{\cal L} & = & \frac{v^2}{4} {\rm Tr} D_\mu \Sigma^\dagger D^\mu \Sigma \left( 1 + 2 a \frac{H}{v} + b \frac{H^2}{v^2} + \dots \right) \nonumber \\
& - & {\bar \psi}^i_L \Sigma \left(1 + c \frac{H}{v} + \dots \right) \nonumber \\
& + & \frac{1}{2} \left(\partial_\mu H \right)^2 + \frac{1}{2} m_H^2 H^2 + d_3 \frac{1}{6} \left(\frac{3 m_H^2}{v} \right) H^3 \nonumber \\
& + & d_4 \frac{1}{24} \left(\frac{3 m_H^2}{v} \right) H^4 + \dots \, ,
\label{calL}
\end{eqnarray}
where
\begin{equation}
\Sigma \; \equiv \; {\rm exp} \left( i \frac{\sigma^a \pi^a}{v} \right) \, .
\label{Sigma}
\end{equation}
The free coefficients $a, b, c, d_3$ and $d_4$ are all normalised so that they are unity
in the SM, and one searches for observable deviations from these values that could be signatures of composite models.

\begin{figure}[htb]
\centering
\includegraphics[height=2.3in]{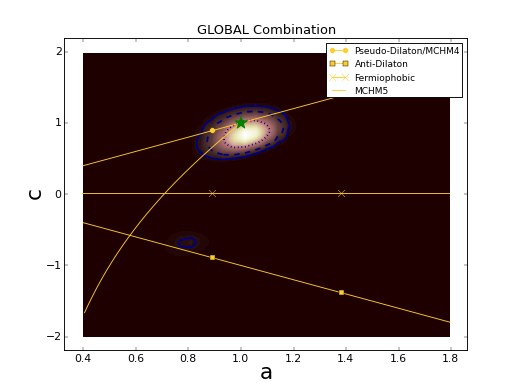}
\caption{\it A global fit to bosonic and fermionic $H$ couplings rescaled by factors $a$ and $c$, respectively.
The SM prediction $a = c = 1$ is shown as the green star~\protect\cite{EY3}, and the yellow lines
show the possible predictions of some composite models.}
\label{fig:EY}
\end{figure}

Fig.~\ref{fig:EY} shows one such analysis~\cite{EY3},
that looked for possible rescalings of the $H$ couplings to bosons by a factor $a$
and to fermions by a factor $c$~\footnote{For a similar recent result from the CMS Collaboration, see~\cite{CMSmH}.
The Higgs Cross Section Working group defines the quantities $\kappa_V \equiv a$ and $\kappa_f \equiv c$~\cite{HiggsxsWG},
which are used by ATLAS and CMS.}.
Fig.~\ref{fig:EY} shows no sign of any deviation from the
SM predictions $a = c = 1$. The yellow lines
in Fig.~\ref{fig:EY} show the predictions of specific composite models that
are excluded unless (in some cases) their predictions are adjusted to
resemble those of the SM.

Since the properties of the Higgs boson as well as other particles
continue to agree with the SM, it is increasingly popular approach to
to regard the SM as an effective field theory (EFT) valid at low energies $< 1$~TeV. The effects of BSM
physics at higher scales may then be parametrised via higher-dimensional EFT operators constructed out of SM
fields,  as a first approximation, with coefficients that can be constrained by precision electroweak data, Higgs data
and triple-gauge couplings (TGCs). 

\begin{figure}[h!]
\centering 
\includegraphics[scale=0.38]{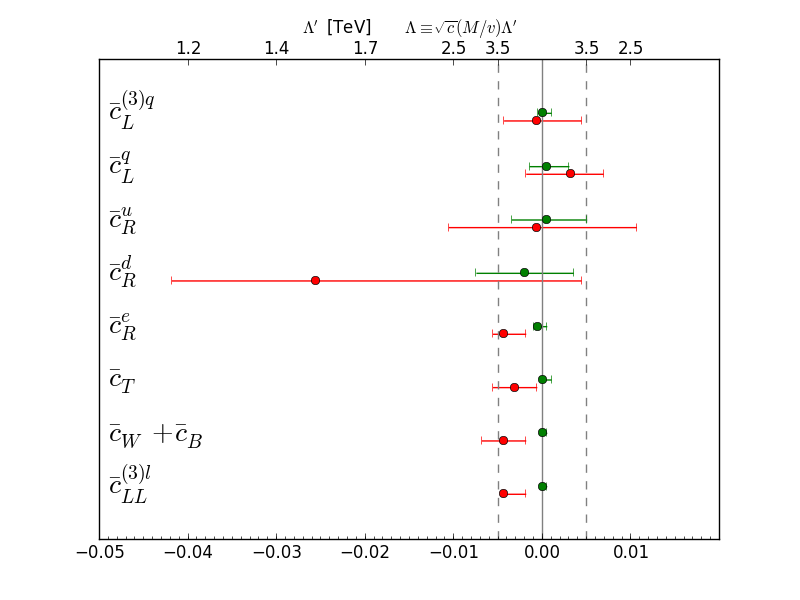}
\caption{\it The 95\% CL ranges found in analyses of the leptonic and hadronic LEP observables. The upper (green)
bars denote single-coefficient fits, and the lower (red) bars denote marginalised multi-coefficient fits.
The upper-axis should be read $\times \frac{m_W}{v}\sim 1/3$ for $\bar{c}_W + \bar{c}_B$.~\protect\cite{ESY4} }
\label{fig:EWPTsummary}
\end{figure}

Ref.~\cite{ESY4} discusses the operators
entering electroweak precision tests (EWPTs) at LEP, together with $95\%$ CL bounds on their individual coefficients
when they are switched on one at a time, and also when marginalised in a simultaneous global fit.
Results for the EFT coefficients $\bar{c}^{(3)l}_{LL}, \bar{c}_T, \bar{c}_W + \bar{c}_B$ and $\bar{c}^e_R$,
which affect the leptonic observables $\{ \Gamma_Z, \sigma^0_{\rm had}, R^0_e, R^0_\mu, R^0_\tau, A^{0,e}_{\rm FB}, m_W \}$,
and the EFT coefficients $\bar{c}^q_L, \bar{c}^{(3)q}_L, \bar{c}^u_R$ and $\bar{c}^d_R$,
which contribute to the hadronic observables $\{R^0_b, R^0_c, A^{0,b}_{\rm FB}, A^{0,c}_{\rm FB}, A_b, A_c\}$,
are shown in Fig.~\ref{fig:EWPTsummary}. 
The upper (green) bars show
the ranges for each of EFT coefficient when it is varied individually, assuming that the other EFT coefficients
vanish, and the lower (red) bars show the ranges for a global fit in which all the EFT coefficients
are allowed to vary simultaneously, neglecting any possible correlations. The ranges of the coefficients
are translated in the legend at the top of the left panel of Fig.~\ref{fig:EWPTsummary}
into ranges of a large mass scale $\Lambda$. All the sensitivities are
in the multi-TeV range. 

\begin{figure}[h!]
\centering 
\includegraphics[scale=0.5]{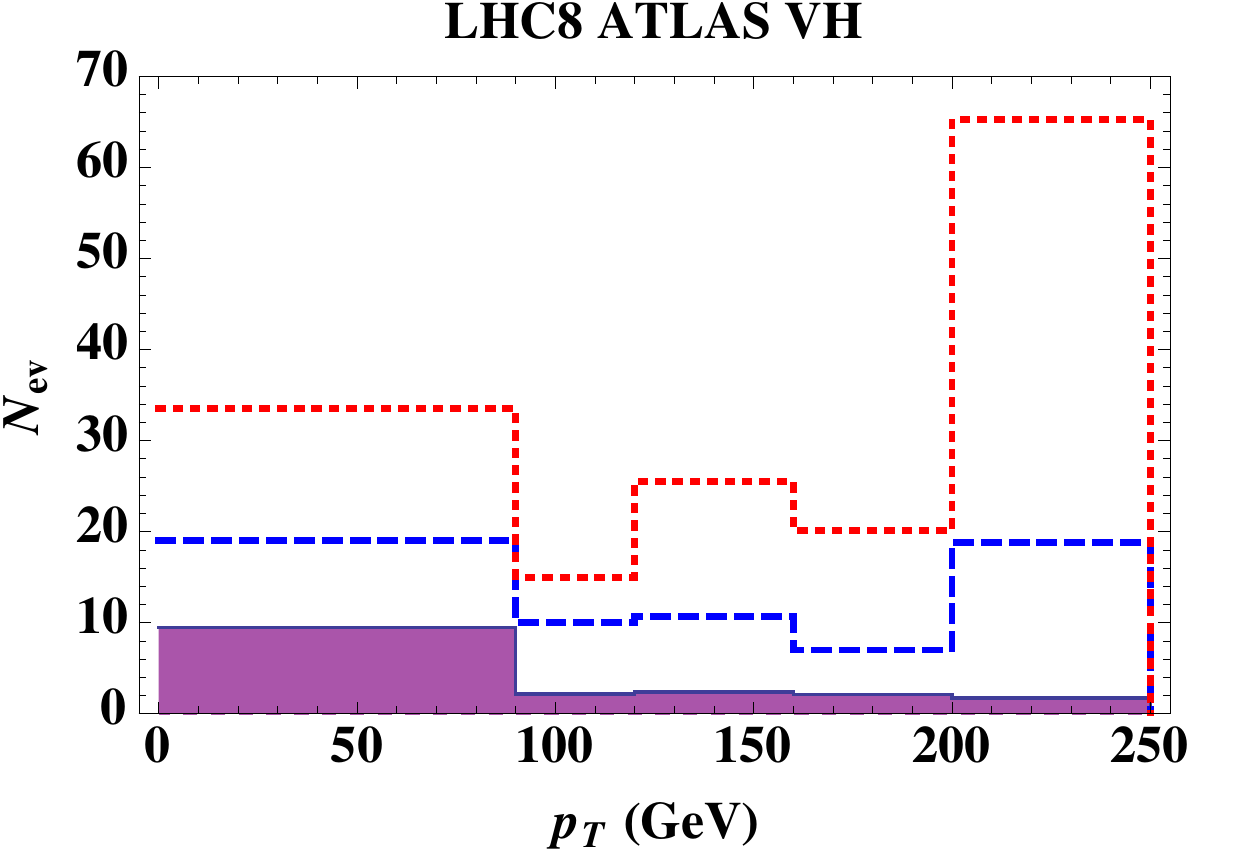} \\
\includegraphics[scale=0.33]{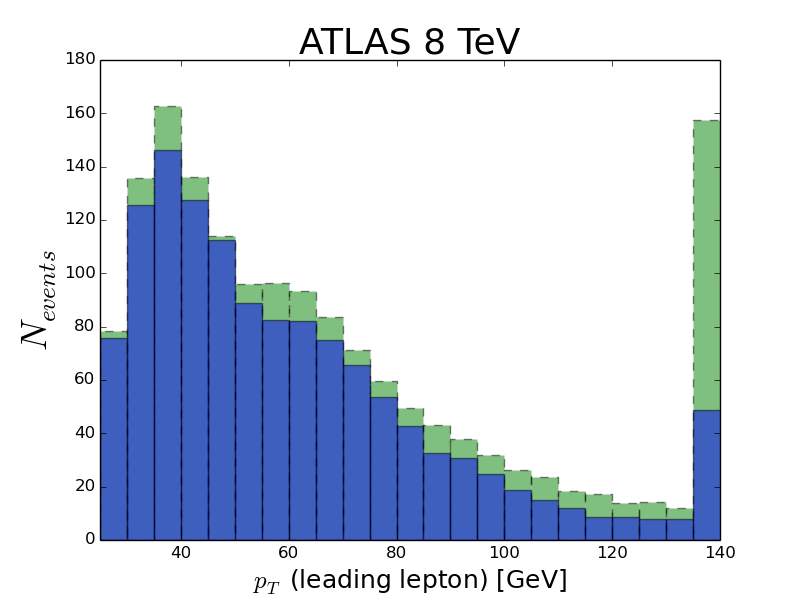}
\caption{\it Upper panel: Simulation of the $p_T^V$ distribution in $(V \to 2 \ell) + (H \to {\bar b}b)$ events at the
LHC showing the SM expectation (purple shading with solid outline), and the distributions with $\bar c_W$ =0.1 and 0.05, respectively
as red-dotted and blue-dashed lines~\protect\cite{ESYHV}.
Lower panel: The same-flavour $p_T$ distribution of the leading lepton in a TGC analysis.
The Standard Model distribution is shaded blue with solid lines,
and the distribution for $\bar{c}_{HW} = 0.1$ is shaded green with dashed lines.
In both cases the last (overflow) bin provides important extra information in addition to the overall normalisation~\protect\cite{ESY4}.}
\label{fig:distributions}
\end{figure}

Other operators contribute to Higgs physics and TGCs, and important information on possible values of their coefficients is
provided by kinematic distributions~\cite{ESYHV}, as well as by total rates, as illustrated in Fig.~\ref{fig:distributions}.

\begin{figure}[h!]
\centering
\includegraphics[scale=0.4]{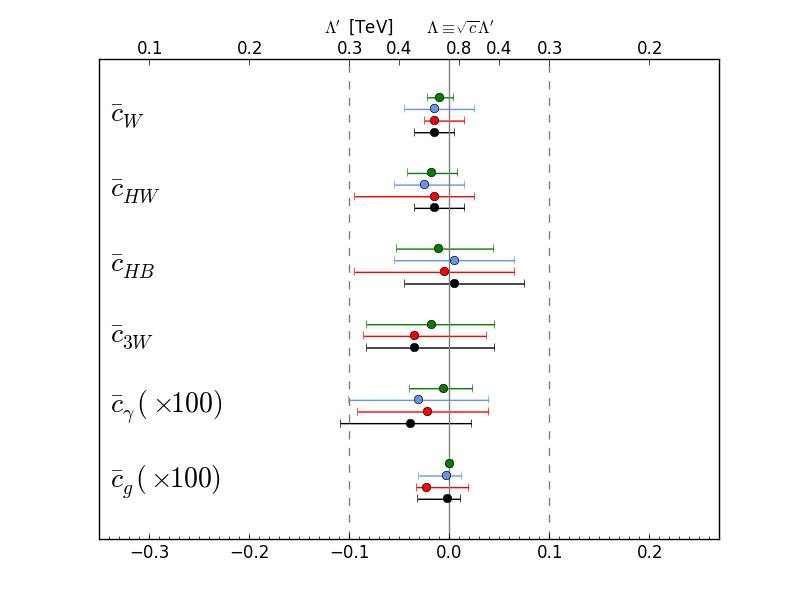}
\caption{\it The 95\% CL constraints for single-coefficient fits (green bars),
and the marginalised 95\% ranges for the
LHC Higgs signal-strength data combined with the kinematic distributions for associated $H + V$ production
measured by ATLAS and D0 (blue bars), with the LHC TGC data (red lines), and the global combination with
both the associated production and TGC data (black bars). Note that $\bar{c}_{\gamma,g}$ are shown $\times 100$,
for which the upper axis should be read $\times 10$~\protect\cite{ESY4}.}
\label{fig:fitsummary}
\end{figure}

Fig.~\ref{fig:fitsummary}~\cite{ESY4} shows a global fit to the Higgs data, including associated production kinematics,
and LHC TGC measurements. The individual 95\% CL constraints with one non-zero EFT coefficient 
at a time are shown as green bars. The other lines show marginalised 95\% ranges fund using the
Higgs signal-strength data in conjunction with the kinematic distributions for associated $H + V$ production
measured by ATLAS and D0 (blue bars), with the LHC TGC data (red lines), and with
both (black bars). The LHC TGC constraints
are the most important for $\bar{c}_{W}$ and $\bar{c}_{3W}$, whereas the Higgs constraints are more
important for $\bar{c}_{HW}$, $\bar{c}_{HB}$ and $\bar{c}_{g}$.

In my view, the EFT approach is the most promising framework for analysing Run~2 results,
in particular because it can be used to tie together many different classes of measurements.

\section{The SM is not enough!}

{\it ``The more important
fundamental laws and facts of physical science have all been discovered"} said Albert Michelson in 1894, just 
before radioactivity and the electron were discovered. {\it ``There is nothing new to be discovered in physics now,
all that remains is more and more precise measurement"} said Lord Kelvin in 1900, just before Einstein's
{\it annus mirabilis} in 1905. Similarly, today there are many reasons to expect physics beyond the SM
even (particularly after the discovery of a (the?) Higgs boson, as I now discuss.

As James Bond might have said~\cite{Bond},
there are 007 important reasons. 1) The measured values of $m_t$ and $m_H$ indicate that
the electroweak vacuum is {\it probably} unstable, in the absence of some BSM physics.
2) The dark matter required by astrophysics and cosmology has no possible origin within the SM.
3) The origin of the matter in the Universe requires additional CP violation beyond CKM.
4) The small neutrino masses seem to require BSM physics.
5) The hierarchy of mass scales could appear more natural in the presence of some new physics at the TeV scale.
6) Cosmological inflation requires BSM physics, since
the effective Higgs potential in the SM would seem to become negative at high scales.
7) Quantising gravity would certainly require physics (far) beyond the SM.

The first two of these issues are discussed in the following.

\section{The Instability of the Electroweak Vacuum}

In the SM with its SU(2)$\times$U(1) symmetry, the origin where $\langle H \rangle = 0$ is unstable and surrounded
by a valley where $\langle H \rangle \equiv v = 246$~GeV, the present electroweak vacuum.
At larger Higgs field values, the effective potential rises, at least for a while.
However, calculations in the SM show that, for the measured values of $m_t$ and $m_H$, the effective potential turns down as a result of 
renormalization of the Higgs self-coupling by the top quark, which overwhelms that by the Higgs itself. Thus, the present electroweak vacuum is
in principle unstable in the SM, and quantum tunnelling generates
collapse into an anti-de-Sitter 'Big Crunch'.

\begin{figure}[htb]
\centering
\includegraphics[height=2in]{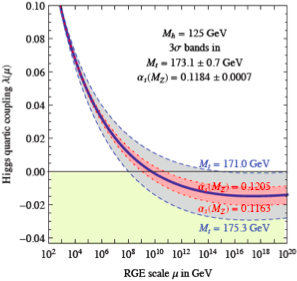} \\
\includegraphics[height=1.6in]{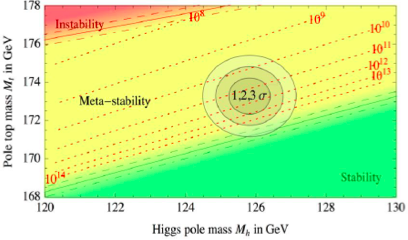}
\caption{\it Left panel: Within the SM, normalisation by the top quark appears to drive the Higgs self-coupling $\lambda < 0$ at
large scales, destabilising the electroweak vacuum.
Right panel: Regions of vacuum stability, metastability and instability in the $(m_H, m_t)$ plane.
Both panels are from~\protect\cite{Buttazzo}.}
\label{fig:Buttazzo}
\end{figure}

The SM calculations~\cite{Buttazzo} shown in the upper panel of Fig.~\ref{fig:Buttazzo}
indicate that the instability sets in at a Higgs scale $\Lambda$:
\begin{eqnarray}
\log_{10} \left( \frac{\Lambda}{{\rm GeV}} \right) & = & 11.3 + 1.0 \left(\frac{m_H}{{\rm GeV}} - 125.66 \right) \nonumber \\
& - & 1.2 \left( \frac{m_t}{{\rm GeV}} - 173.10 \right) \nonumber \\
& + & 0.4 \left(\frac{\alpha_s(M_Z) - 0.1184}{0.0007} \right) \, .
\label{Buttazzo}
\end{eqnarray}
Uisng the world average values of $m_t$, $m_H$ and $\alpha_s (M_Z)$, this formula yields
\begin{equation}
\Lambda \; = \; 10^{10.5 \pm 1.1}~{\rm GeV} \, .
\label{Lambda}
\end{equation}
However, we see in the lower panel of Fig.~\ref{fig:Buttazzo} that this calculation is very sensitive to $m_t$.
Note in this connection that the D0 Collaboration has recently reported a new, higher, value of $m_t$~\cite{D0mt}
(tending to make the vacuum more unstable),
whereas the CMS Collaboration has
reported lower values of $m_t$ from new analyses~\cite{CMSmt} (tending to make
the vacuum more stable). A more accurate experimental measurement of $m_t$ would
help us understand the fate of the Universe within the SM, and the possible need for BSM physics
to stabilise the electroweak vacuum, but we also need to understand better the relationship
between the parameter $m_t$ in the SM Lagrangian and the effective mass parameter
measured by experiments~\cite{Moch}.

Even if the lifetime of the electroweak vacuum
is much longer than the age of the Universe, as suggested by these calculations,
one cannot simply ignore this problem.
The early Universe is thought to have had a very
high energy density, e.g., during an inflationary epoch~\cite{CMB}, at which time quantum and thermal
fluctuations at that time would have populated
the anti-de-Sitter `Big Crunch' region~\cite{Oops}. However, it is possible that we
might have been lucky enough to live in a non-anti-de-Sitter region and thereby surviving~\cite{maybenot}. The
problem could be avoided with suitable new physics beyond the SM, of which
one example is supersymmetry~\cite{ER}.

\section{Supersymmetry}

One may love supersymmetry (SUSY) for many reasons, such as 
rendering the hierarchy problem more natural, providing
a candidate for the cold dark matter, aiding grand unification
and its essential (?) r\^ole in string theory. In my mind, Run~1 of the LHC has added
three more reasons, namely the mass of the Higgs boson, which was
predicted successfully by supersymmetry~\cite{SUSYmH,FH}, the fact that the Higgs couplings are similar to
those of the SM Higgs boson, as discussed earlier and as expected in simple realisations of the MSSM~\cite{EHOW},
and the stabilisation of the electroweak vacuum, as mentioned just above. How can we resist SUSY's manifold charms?

However, so far SUSY has kept coyly out of sight in searches at the LHC, direct searches for the scattering of dark matter particles,
indirect searches in flavour physics, etc.. Where could SUSY
be hiding? We know that SUSY must be a broken symmetry, but we do not know how, so we do not know
what the SUSY spectrum may be. It is often assumed that there is a discrete R-symmetry, which
would guarantee the stability of the lightest supersymmetric particle (LSP), providing the above-mentioned
dark matter candidate. It is often assumed that the SUSY-breaking sparticle masses are universal at some high
renormalisation scale, usually the GUT scale, but this has no strong theoretical justification. The simplest model is one in which all the SUSY-breaking
contributions $m_0$ to the squark, slepton and Higgs masses are equal at the GUT scale, and
the SU(3), SU(2) and U(1) gauging masses $m_{1/2}$ are also universal, which is called
the constrained MSSM (CMSSM). It could also be that the SUSY-breaking
contributions to the masses of the two Higgs doublets of the MSSM differ from
those of the squarks and leptons, and may be equal to each other (the NUHM1), or different from each other (the NUHM2).
Alternatively, one may consider the phenomenological MSSM (pMSSM) in which no GUT-scale universality is assumed. 

\begin{figure}[htb]
\resizebox{7cm}{!}{\includegraphics{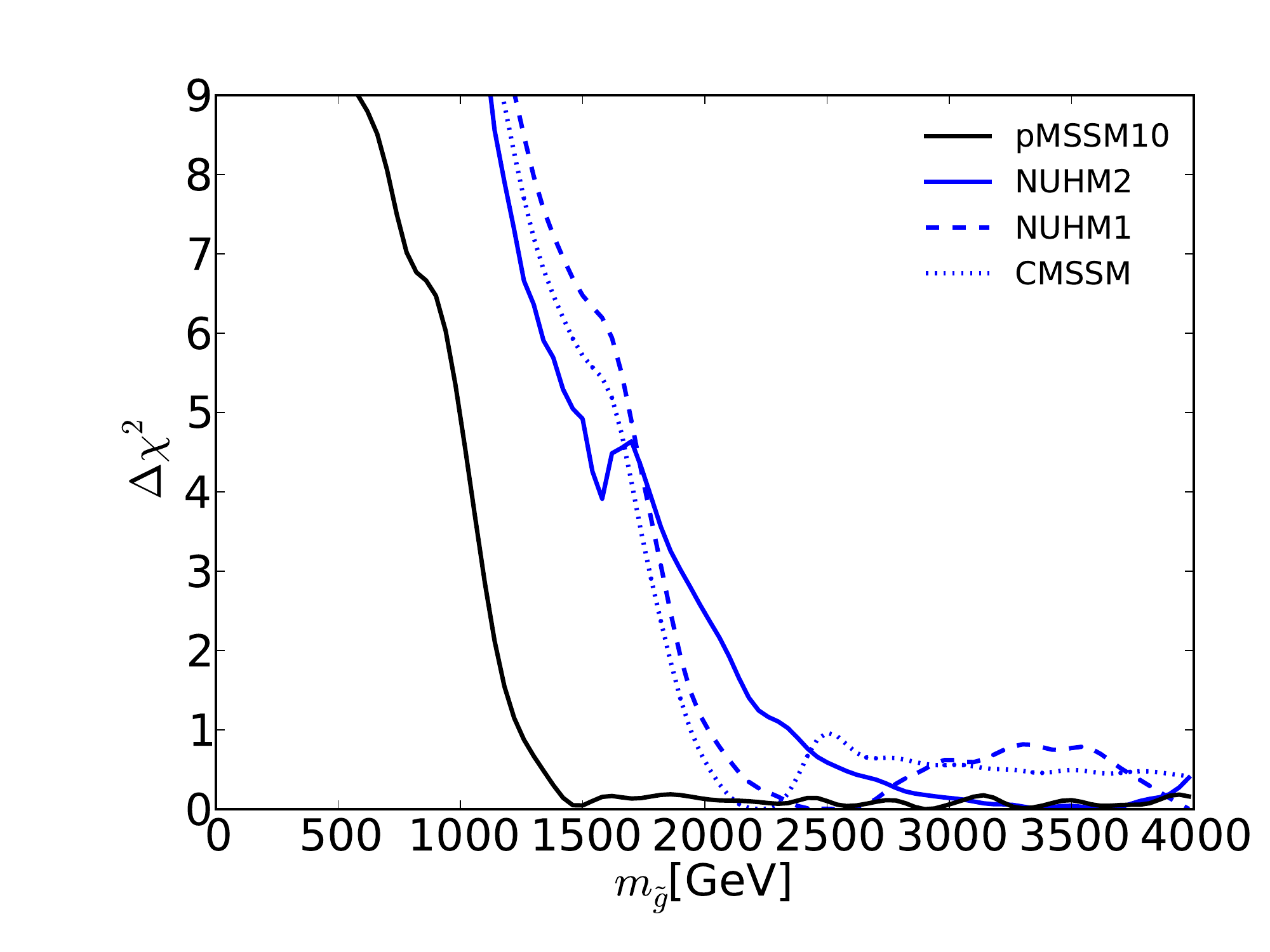}} \\
\resizebox{7cm}{!}{\includegraphics{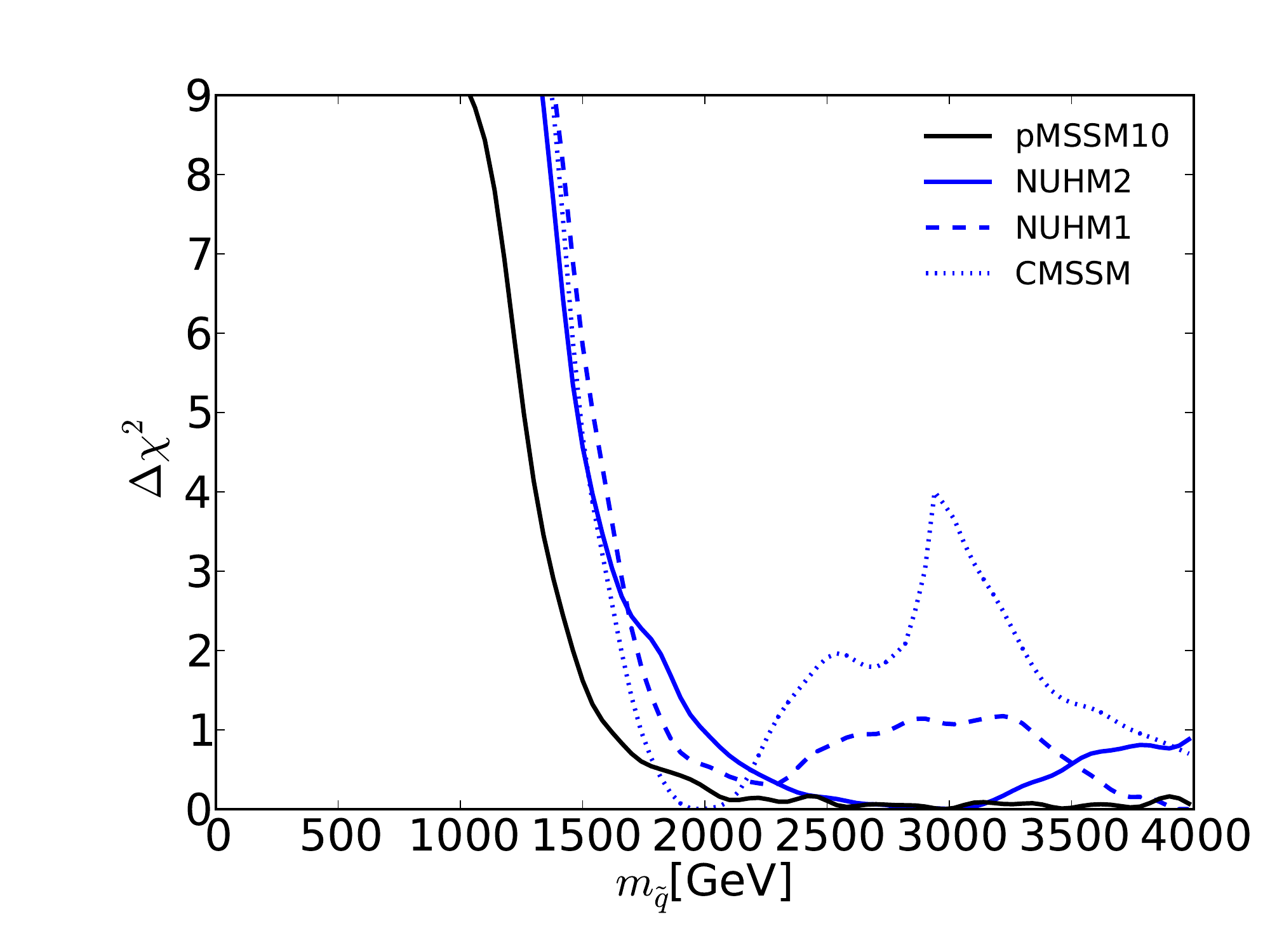}}
\caption{\it The profile likelihood functions
for the gluino mass (upper panel) and the first- and second-generation squark masses (lower panel). 
The solid black lineis are for the pMSSM10~\protect\cite{MC11}, the solid blue lines for the NUHM2~\protect\cite{MC10},
the dashed blue lines for the NUHM1 and the dotted blue lines for the
CMSSM~\protect\cite{MC9}.}
\label{fig:MC11}
\end{figure}

Some results of global fits to the CMSSM, NUHM1, NUHM2 and a version of the pMSSM with 10 free SUSY-breaking parameters,
combining all experimental and phenomenological constraints and requiring that the relic supersymmetric particle density be
within the cosmological range, are shown in Fig.~\ref{fig:MC11}~\cite{MC9,MC10,MC11}. The upper panel shows
the profile likelihood functions for the gluino mass in these models, and the lower panel shows the likelihood functions
for the first-and second-generation squarks (which are assumed to be equal in the pMSSM10). In the GUT-universal models the 95\% CL
lower limits on the squark and gluino masses are $\sim 1.5$~GeV, whereas they could be significantly lighter in the pMSSM10,
which offers greater prospects for discovering SUSY in LHC Run~2~\cite{MC11}.

\begin{figure}[htb]
\centering
\resizebox{7cm}{!}{\includegraphics{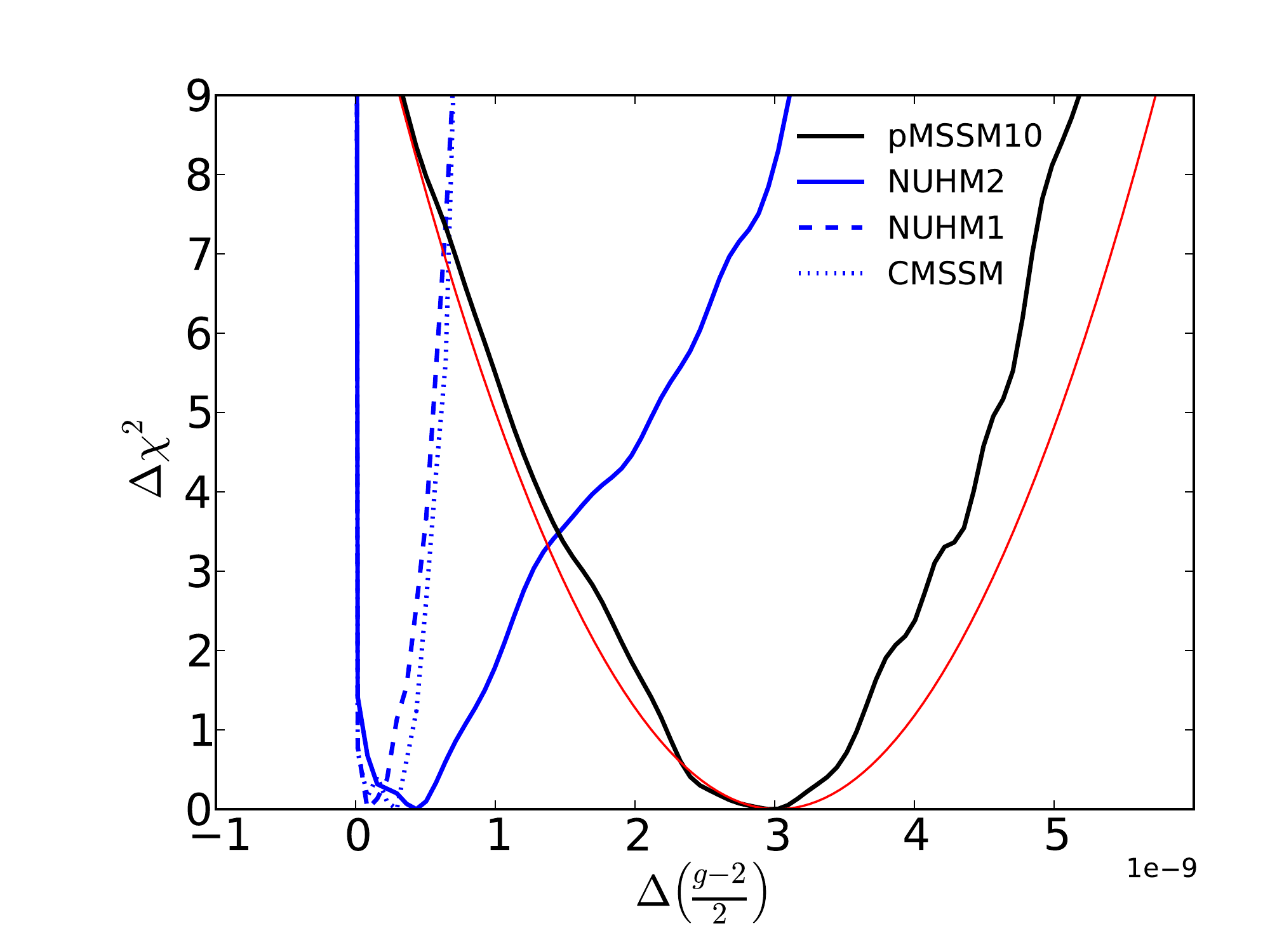}}
\caption{\it The one-dimensional $\chi^2$ likelihood function for $g_\mu - 2$
in the CMSSM, NUHM1, NUHM2 (blue lines) and the pMSSM10 (black line)~\protect\cite{MC11}. The red line
represents the uncertainty in the experimental range of $g_\mu - 2$.}
\label{fig:g-2}
\end{figure}

The pMSSM revives the possibility of explaining the discrepancy between the
SM calculation of $g_\mu - 2$ and the experimental measurement within a SUSY model. This is not possible in
the CMSSM, NUHM1 and NUHM2, because of the LHC constraints, and these models
predict values of the $g_\mu - 2$ similar to the SM prediction,
as shown by the blue lines in Fig.~\ref{fig:g-2}. However, the black line in this Figure shows that the experimental
measurement could be accommodated within the pMSSM~\cite{MC11}.
There are plans for two new experiments to measure $g_\mu - 2$~\cite{futureg-2},
and other low-energy $e^+ e^-$ experiments will help clarify the discrepancy between
the SM and experiment. 

If this is indeed due to SUSY, our pMSSM10
analysis suggests that its discovery may not be far away!
In particular, there are prospects in searches for jets + missing transverse energy searches
at the LHC, as well as dedicated searches for sleptons and light stop squarks~\cite{MC11}.

\section{Dark Matter Searches}

\begin{figure}[htb]
\centering
\resizebox{6cm}{!}{\includegraphics{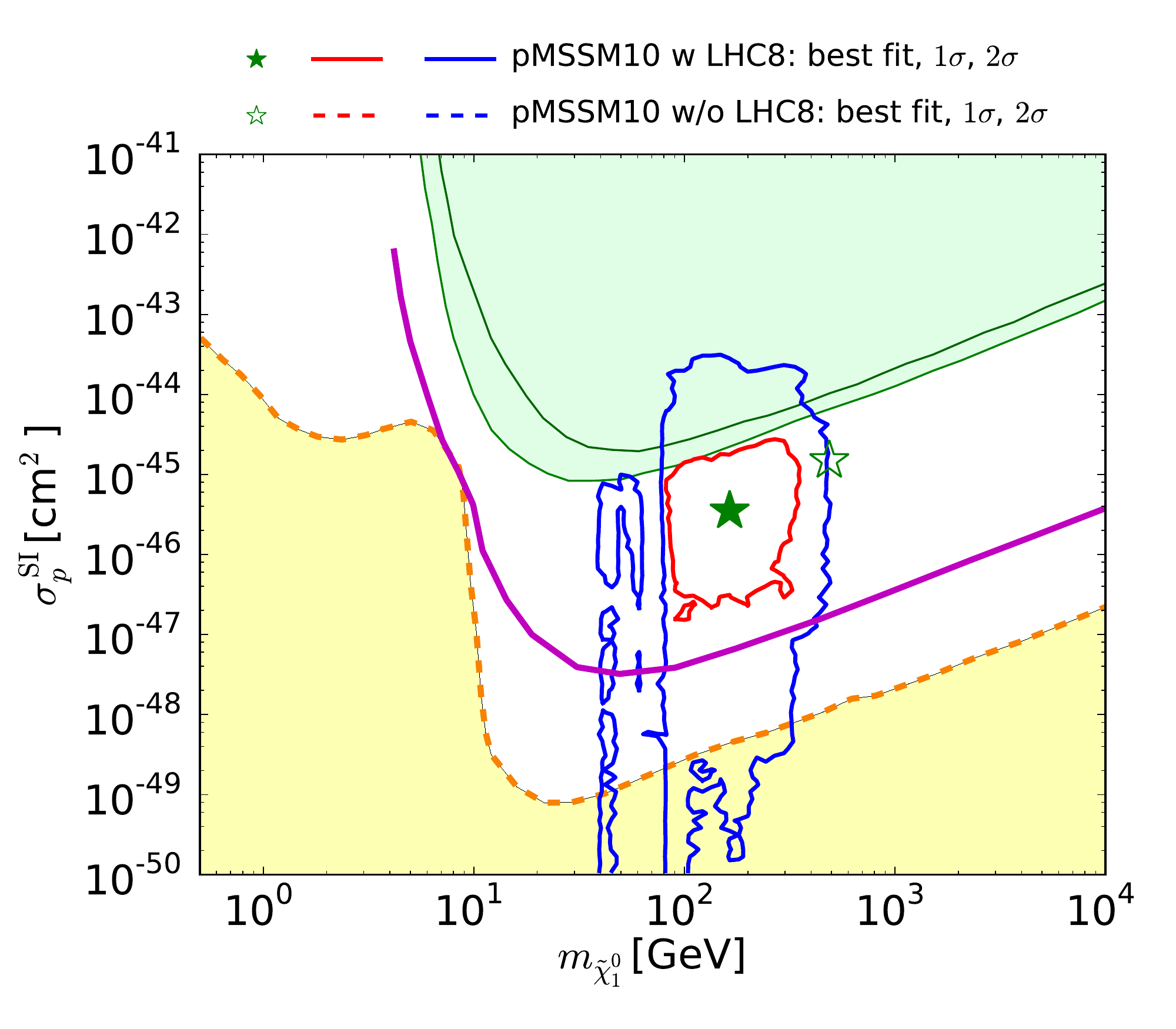}} \\
\includegraphics[height=1.4in]{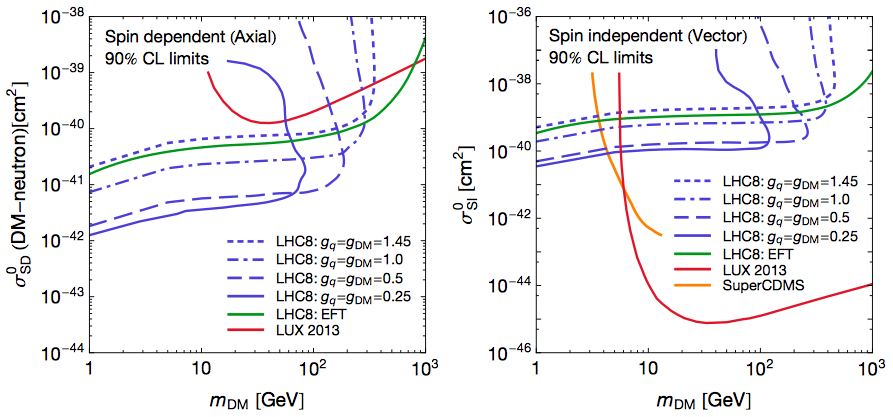}
\caption{\it Upper panel: The two-dimensional
profile likelihood function in the pMSSM10 in the $(m_{\tilde \chi_0^1}, \sigma^{\rm SI}_p)$-plane~\protect\cite{MC11}, showing the regions
excluded by the XENON100 and LUX experiments (shaded green), the neutrino `floor'
(shaded yellow), and the prospective sensitivity of the LZ experiment~\protect\cite{LZ}. The preferred 68\% CL region is outlined in red,
and the region allowed at the 95\% CL is outlined in blue. The solid contours incorporate the LHC constraints, and the dashed contours omit them.
Lower panels: A comparison of the current 90\% CL direct search limits from LUX and SuperCDMS
(red and orange lines, respectively), the monojet limits in simple models (blue lines) and the limits in
an effective field theory framework (green line) in the cross section vs $m_{DM}$ plane used by the direct detection community.
The left and right panels show, respectively, the limits on the spin-dependent and spin-independent cross sections
appropriate for axial- vector and vector mediators~\protect\cite{ICDM}. }
\label{fig:DMcomparison}
\end{figure}

As already mentioned, a supersymmetric model that conserves R-parity has a natural
candidate for a cold dark matter particle, and this is often taken to be the lightest neutralino
${\tilde \chi_0^1}$~\cite{EGNOS} (though other candidates are also possible). The present limits from direct searches for the scattering of massive
cold dark matter particles in underground experiments are shown in the upper panel of Fig.~\ref{fig:DMcomparison}, together
with predictions in the pMSSM10~\cite{MC11}. The 68\% CL region in this model (outlined by the red contour) lies just below the
current experimental limit and within range of the planned LZ experiment (magenta line)~\cite{LZ}. 

Other TeV-scale extensions
of the SM, such as some extra-dimensional models with K-parity and little Higgs models with T-parity,
also have possible candidates. It is therefore useful to make a model-independent comparison of the capabilities
of the LHC and direct dark matter searches, and one such is shown in
the lower panel of Fig.~\ref{fig:DMcomparison}. This compares direct
astrophysical searches for the scattering of generic TeV-scale dark matter particles with the current reaches of the LHC via monojet searches,
for the cases of spin-dependent (axial) couplings (left panel) and spin-independent (vector) couplings
(right panel)~\cite{ICDM}. In the former case the LHC monojet searches generally have greater sensitivity
than the direct searches, except for dark matter particle masses $\gtrsim 1$~TeV where the LHC
runs out of phase space. On the other hand, direct searches for spin-independent interactions are
more sensitive than the LHC searches for masses $\gtrsim 4$~GeV. SUSY models generally
favour a relatively large mass for the dark matter particle, the pMSSM10 being one example, as seen in the upper panel of Fig.~\ref{fig:DMcomparison}.

\section{Possible Future Colliders}

\begin{figure}[htb]
\centering
\includegraphics[height=2in]{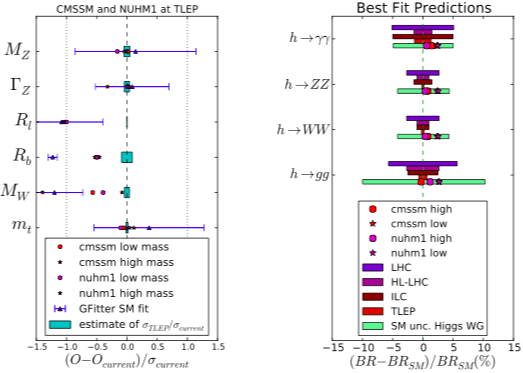}
\caption{\it Comparison of the present precisions in measurements of various $Z$ properties (left panel) and Higgs couplings
(right panel), together with the prospective precisions of possible measurements at future colliders
and the deviations from the SM predictions found at the best-fit points in various SUSY models. The right panel
also shows the current theoretical uncertainties. From~\protect\cite{TLEP}.}
\label{fig:TLEP}
\end{figure}

There is longstanding interest in building
a high-energy $e^+ e^-$ collider, which might be linear such as the ILC ($E_{CM} \lesssim 1$~TeV) or CLIC ($E_{CM} \lesssim 3$~TeV).
On the other hand, there is now interest in Europe and China in a possible large circular tunnel to contain an
$e^+ e^-$ collider with $E_{CM} \lesssim 350$~GeV and/or a $pp$ collider with $E_{CM} \lesssim 100$~TeV~\cite{FCC}. A circular
$e^+ e^-$ collider would provide measurements of the $Z$ and Higgs bosons of unparalleled accuracy,
as seen in Fig.~\ref{fig:TLEP}~\cite{TLEP}. These would be able to distinguish between the predictions of
the SM and various fits in the CMSSM, NUHM1 and NUHM2, as shown, if one can also
reduce correspondingly the present theoretical uncertainties, which are indicated in the right panel by the shaded green bars.
The other coloured bars illustrate the accuracies attainable with measurements at various accelerators.

\begin{figure}[!t]
 \vspace{0.5cm}
 \begin{center}
  \includegraphics[width=6.5cm]{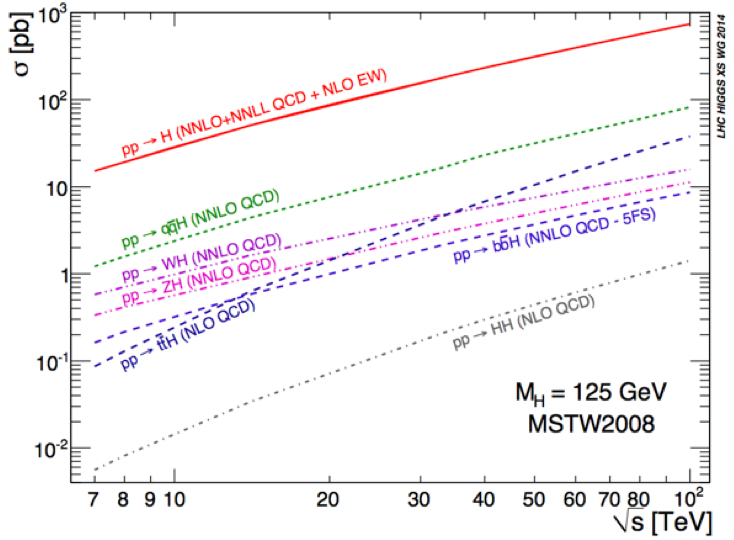} \\
   \vspace{0.5cm}
  \hspace{0.3cm}
  \includegraphics[width=6.7cm]{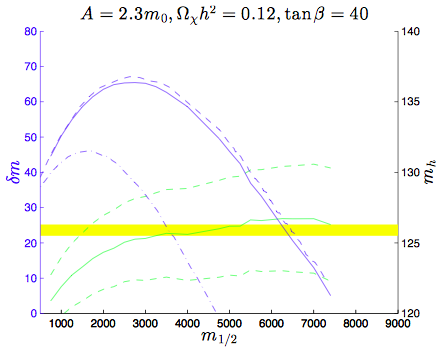}
\caption{\it Upper panel: Cross sections for various Higgs production processes at $pp$ colliders
as functions of the centre-of-mass energy~\protect\cite{FCC-hh}. Lower panel: One of the possibilities for a relatively
heavy SUSY dark matter particle weighing $\sim 0.4 m_{1/2} \lesssim 3$~TeV. The
vertical axis is the mass difference between the dark matter particle and the next-to-lightest
supersymmetric particle, in this case the lighter stop squark. The solid and upper dashed blue lines
correspond to the current central and +1$\sigma$ values of the dark matter density, the horizontal
yellow band represents the experimental value of the Higgs mass, and the green solid and dashed
lines represent the central value and $\pm$1$\sigma$ uncertainties in
theoretical calculations of the Higgs mass~\protect\cite{EOZ}.}
\label{fig:FCC-hh}
 \end{center}
\end{figure}

A future high-energy $pp$ collider would produce many more
Higgs bosons than the LHC, as seen in the upper panel of Fig.~\ref{fig:FCC-hh}~\cite{FCC-hh}, offering
the possibility of measuring Higgs couplings with greater statistical accuracy, and also including the
elusive triple-Higgs coupling. A high-energy $pp$ collider would also offer unique possibilities
to discover and/or measure the properties of SUSY particles. Even the SUSY dark matter particle
could weigh several TeV, as seen in the lower panel of Fig.~\ref{fig:FCC-hh}~\cite{EOZ},
which illustrates a strip in the CMSSM parameter space where the relic neutralino density is
brought into the the range allowed by cosmology through coannihilation with the lighter
stop squark. In the example shown, the lightest neutralino weighs $\lesssim 3$~TeV and
only a $pp$ collider with $E_{CM} \sim 100$~TeV would be able to explore all the range of
particle masses compatible with SUSY providing dark matter (solid
and upper dashed blue lines). For all this range calculations of the
Higgs mass are compatible with the experimental value (represented by the yellow band),
considering the theoretical uncertainties represented by the solid and dashed green lines.

\begin{figure}[htb]
\centering
\includegraphics[height=2.3in]{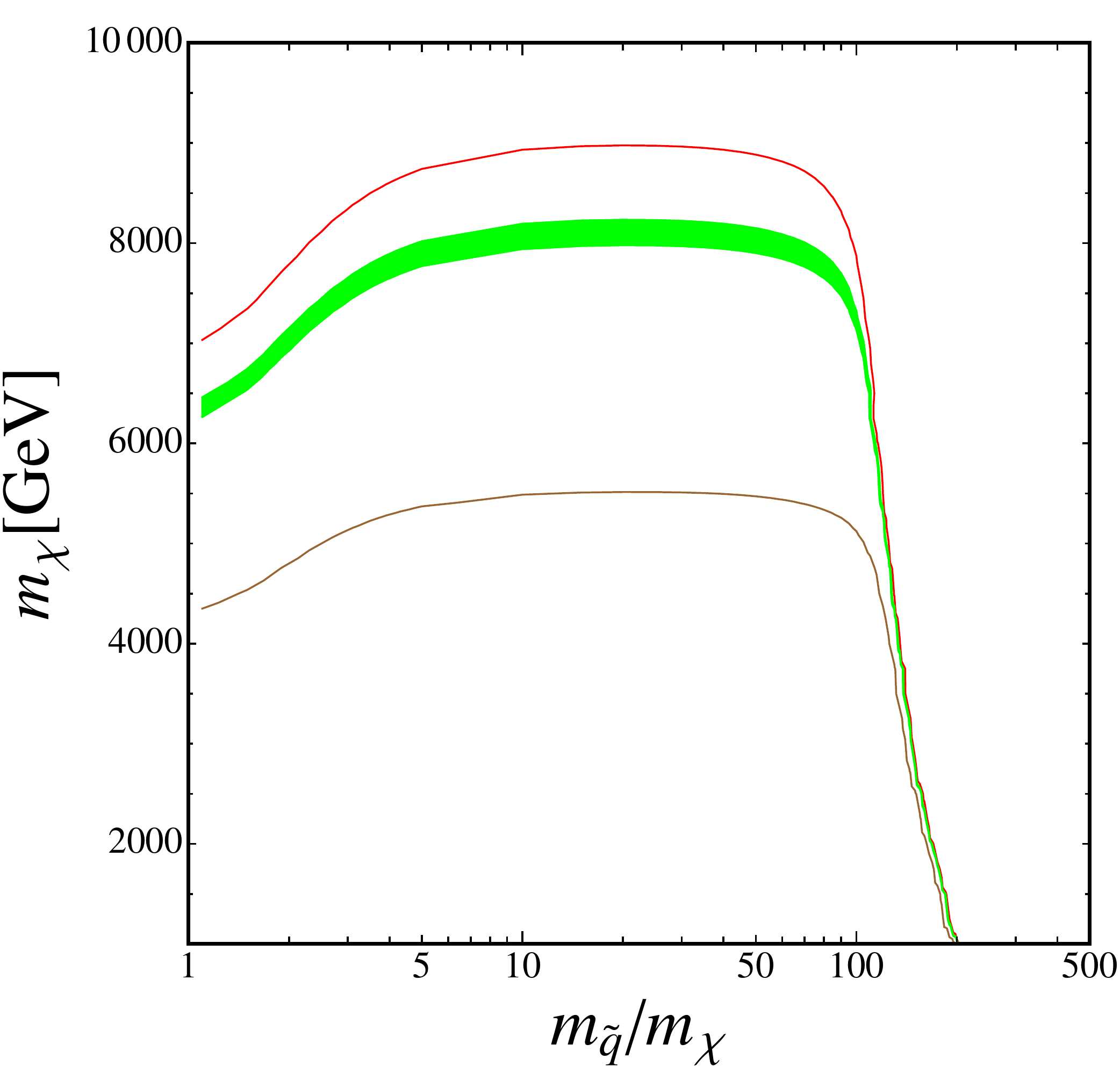}
\caption{\it The mass of the dark matter particle (assumed to be a Bino) at the endpoint 
of the gluino coannihilation strip when $\Delta m = m_{\tilde g} - m_\chi = 0$,
as a function of $m_{\tilde q}/m_{\tilde g}$. The green band corresponds to the current $3$-$\sigma$ range of
the dark matter density: $\Omega_\chi h^2 = 0.1193 \pm 0.0042$, and the brown and red contours are for $\Omega_\chi h^2 = 0.05$ and $0.15$,
respectively. From~\protect\cite{ELO}.}
\label{fig:ELO}
\end{figure}

The supersymmetric dark matter particle might be even heavier in more general supersymmetric
models. For example, if the lightest neutralino coannihilates with an almost degenerate gluino, it
may weigh $\lesssim 8$~TeV, as seen in Fig.~\ref{fig:ELO}, which would be a challenge even
for a 100-TeV collider.

The physics cases for future large circular colliders are still being explored. There will be
bread-and-butter high-precision Higgs and other SM measurements to probe possible BSM
scenarios for physics. As for direct searches for new physics, the search
for dark matter particles may provide the strongest case, and this is under continuing
study. 

\section{Conclusion}

The physics landscape will look completely different when/if future
runs of the LHC find evidence for new physics beyond
the SM such as SUSY. The LHC adventure has only just begun, and we look forward to a big increase in energy with Run~2 and
eventually two orders of magnitude more integrated luminosity.  Lovers of SUSY should not be disappointed that she has not yet appeared.
It took 48 years for the Higgs boson to be discovered, but
four-dimensional SUSY models were first written down just 41 years ago~\cite{WZ}.
We can be patient
for a while longer.

\section*{Acknowledgements}
The author is supported in part by
the London Centre for Terauniverse Studies (LCTS), using funding from
the European Research Council 
via the Advanced Investigator Grant 267352, and in part by STFC
(UK) via the research grants ST/J002798/1 and ST/L000326/1.

\end{document}